\documentclass[journal, a4paper, twoside, twocolumn, 9pt]{IEEEtran}
\usepackage{float}
\usepackage{geometry}
\usepackage{amsmath, amssymb, mathtools}
\usepackage{graphicx}
\usepackage[american, siunitx, nooldvoltagedirection, americanresistors, cuteinductors]{circuitikz}
\usetikzlibrary{shapes.geometric}
\usetikzlibrary{positioning}
\tikzset{
    block/.style = {draw, rectangle, minimum height=2em, minimum width=3em},
    sum/.style = {draw, circle, inner sep=0pt, minimum size=0.8cm, node distance=1cm},
    input/.style = {coordinate},
    output/.style = {coordinate}
}
\usepackage{lipsum} % Only for testing; remove in final version
\usepackage[utf8]{inputenc}
\usepackage[T1]{fontenc}
\usepackage{lmodern}
\usepackage[italian,english]{babel}
\usepackage{upgreek}
\usepackage{subfig}
\usepackage{longtable}
\usepackage{booktabs}
\usepackage{multirow}
\usepackage{hyperref}

\newcommand{\bx}{{\bf x}}
\newcommand{\bK}{{\bf K}}
\newcommand{\bA}{{\bf A}}
\newcommand{\bB}{{\bf B}}
\newcommand{\bC}{{\bf C}}
\newcommand{\bw}{{\bf w}}
\newcommand*{\pauline}{\textcolor{black}}
\begin{document}
 % Required for clickable links

\title{\textbf{\huge{Data-Driven Control Of Power Converters}}} 
\author{
   Marwan Soliman$^1$, Pauline Kergus$^1$, Diego Regruto$^3$, Luiz Villa$^2$, Zohra Kader$^1$ 
   
   \fontsize{10pt}{10pt}\selectfont
   $^1$ LAPLACE, Université de Toulouse, CNRS, INPT, UPS, Toulouse, France.
   
   $^2$ LAAS, Université de Toulouse, CNRS, Toulouse, France.
   
   $^3$ Politecnico di Torino, DAUIN Dept, Turin, Italy.
   
   \small Corresponding \href{mailto:Soliman@laplace.univ-tlse.fr}{Soliman@laplace.univ-tlse.fr}
}
\IEEEtitleabstractindextext{%
    \begin{abstract} 
    The fundamental role of power converters is to efficiently manage and control  the flow of electrical energy, ensuring compatibility between power sources and loads.
All these applications of power converters need the design of an appropriate control law. Control of power converters is a challenging problem due to the presence of switching devices which are difficult to handle using traditional control approaches.
The objective of this paper is to investigate the use of data-driven techniques, in particular the Virtual References Feedback Tuning (VRFT) method, in the context of power converters feedback control. This study considers a buck \pauline{mode} power converter circuit provided by the OwnTech foundation.

	\end{abstract}}
 \maketitle										% Stampa il titolo
\selectlanguage{english}

\IEEEdisplaynontitleabstractindextext		
\vspace{-3mm}
\section{Introduction}  
In recent years, the field of power electronics has witnessed significant advancements, not only in the design of power converters but also in the development of their control strategies. Traditional control \pauline{of power converters} typically rely on the derivation of an accurate mathematical model of the system, followed by linearization, after which a controller is designed based on the resulting simplified dynamics \cite{ref18}. However, obtaining precise models for power converters \pauline{and applying this control pipeline may become challenging for complex converter architecture}. \pauline{In addition, due to their inherent nonlinearities, fast switching behavior, parameters' uncertainties, and operating condition variability, the control performances and the stability cannot be guaranteed by following this traditional method}. \pauline{Indeed}, the linearization process limits the validity of the controller to a narrow operating range, reducing overall system performance in practical applications.

\pauline{To sum up, the main challenge when designing a control law for a power converter boils down to the obtention of a control-oriented model, simple yet accurate enough to use model-based control techniques.} Data-driven control techniques have emerged as a promising alternative: \pauline{they allow to bypass this complex modelling phase and to use data directly to tune a controller. \pauline{These techniques gained attention in the power electronics community lately, see \cite{remes2020virtual,ref34} for instance.} An overview of data-driven control techniques can be found in \cite{hou2013model} and \cite{markovsky2023data}.} Among them, Virtual Reference Feedback Tuning (VRFT) \pauline{\cite{ref5}} enables controller design directly from input-output data, eliminating the need for detailed modeling. By relying exclusively on experimental or simulated data, VRFT facilitates rapid controller \pauline{synthesis}, making it particularly suitable for power converter applications. \pauline{The closest work to this paper are \cite{remes2020virtual} and \cite{ref34}. In \cite{remes2020virtual}, a variant of VRFT based on a flexible reference model to be achieved by the controlled system is used. In comparison, the present paper focuses on a VRFT variant that includes an anti-windup compensation \cite{breschi2020direct}. Such anti-windup compensation is also considered in \cite{ref34} on a current mode buck converter.}

%The adoption of VRFT for power converter control offers several benefits, including reduced dependency on accurate system models, faster controller deployment, and enhanced robustness to model uncertainties. These advantages align with the growing demand for efficient and reliable control solutions in applications such as renewable energy systems, electric vehicles, and smart grids.

This study focuses on the development and implementation of a data-driven \pauline{controller using the VRFT methodology} for a buck-mode power converter, \pauline{which Simulink model is} provided by the OwnTech Foundation \cite{ref21}. %To illustrate the complexities inherent to traditional model-based approaches, the derivation of the converter’s state-space averaged model is also presented.
This paper is organized as follows: Section 2 \pauline{introduces the considered} power converter \pauline{and the associated averaged state-space model}. %, and presents a comparison between the real system and the SSA model to validate its accuracy.
Section 3 details \pauline{the data-driven tuning of a PID controller for the considered converter through VRFT, with or without anti-windup.} %Section 3 addresses the windup phenomenon and introduces an extended VRFT technique with the anti-windup method to prevent the integrator windup phenomenon. 
Additionally, section 3 \pauline{provides simulation results of the resulting closed-loops}. Finally, \pauline{conclusions and outlooks are given} in Section 4.\\

%\textbf{Notations:}\\ 
%The symbol $\otimes$ represents the Kronecker product between two matrices. If A is an $n\times m$ matrix and B is an $p\times q$ matrix, then the kronecker product $A\otimes B$ is $np \times mq$ matrix: 
% \[
%   A\otimes B=\begin{bmatrix}
%   a_{11}B& \hdots &a_{1m}B\\
%   \vdots & \ddots & \vdots \\
%   a_{n1}B & \hdots & a_{nm}B
% \end{bmatrix}.
% \]

\section{\pauline{Considered DC-DC Converter}}
\par
\begin{figure*}[htbp]
    \centering
    \includegraphics[width=\textwidth,trim={0 22cm 0 0},clip]{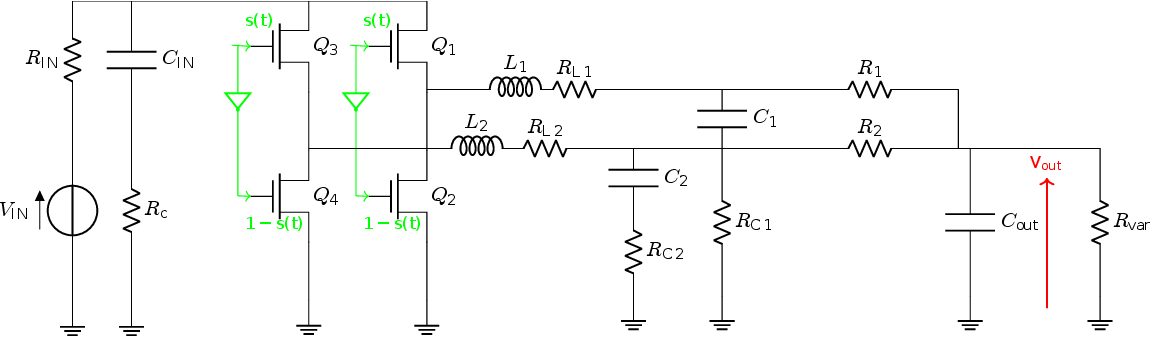}
    \caption{OwnTech's converter circuit.}
    \label{fig:fullwidth}
\end{figure*}

\begin{figure*}[htbp]
\centering
\includegraphics[width=\textwidth]{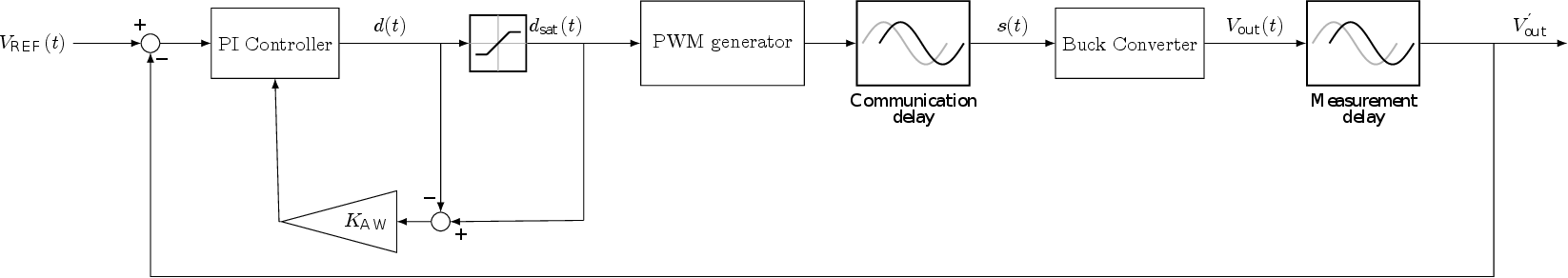}
\caption{Considered closed-loop control structure.}
\label{fig:control_structure}
\end{figure*}

\begin{figure} [h]
    \centering
    \includegraphics[width=0.8\linewidth]{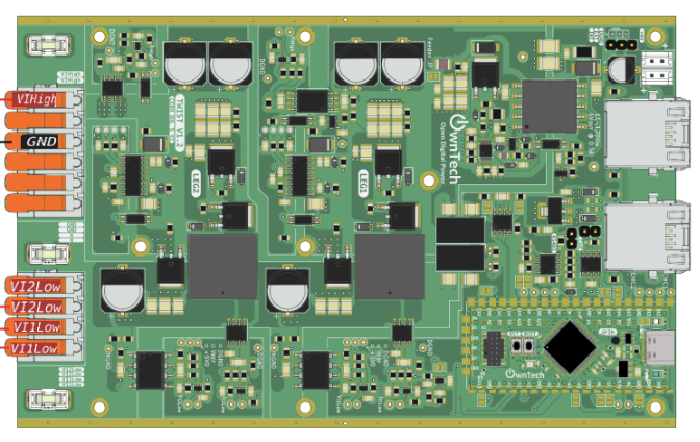}
    \caption{OwnTech's power converter.}
    \label{fig:333}
\end{figure}

 \pauline{A voltage mode buck converter from the OwnTech Foundation \cite{ref21}, see Figure \ref{fig:333}, is considered in this work. The corresponding circuit is given in Figure \ref{fig:fullwidth}, and is implemented in Simulink (the model is publicly available at \cite{ref21}). It consists of three main parts:}
\begin{itemize} 
    \item \textbf{The converter circuit:} A buck mode converter consisting of two parallel legs, \pauline{as shown on Figure \ref{fig:fullwidth}}. 
    \item \textbf{The voltage input source:} A simple DC-DC battery with an input voltage $V_{IN}$ and an internal resistance $R_{IN}$.
    \item \textbf{The load:} A droop bus load with two resistances, a capacitor $C_{out}$ and a variable resistance $R_{var}=R_0+\delta_{R}$. The load variation is treated as a disturbance in this work.  
\end{itemize}
\pauline{The circuit parameters are given in Table \ref{tab:measurements}. The whole controlled system is visible on Figure \ref{fig:control_structure}. Both legs are controlled through a PWM signal $d$. The objective is to regulate the output voltage around a desired value $V_{REF}$. To do so, an output feedback controller is designed using a data-driven control technique in the next section. The simulation model includes a PWM generator model : the switching period is denoted $T_s$ and the sampling period of the whole Simulink model is denoted $T_{samp}$. The PWM generator has a higher time resolution ${\Delta t}_{PWM}$. This model also accounts for communication delays $\tau_{\text{comm}}$ and measurements delays $\tau_{\text{meas}}$.}

%\textcolor{red}{What is the PWM period value? What's the sampling period you use for d2c of the VRFT PI?}
% Delta t indicate the PWM resolution 

% Figure \ref{figure: Timer model} shows the timer model used to generate the switching signals. It receives the duty cycle $d(t)$ as input ,which is sampled by a Zero-Order Hold block, limited by a saturation block, and then processed by a PWM generator with phase delay. The output is the switching signal $s(t)$
\begin{table}[h]
    \centering
    \caption{\pauline{Circuit parameters of the OwnTech converter model}}
    \normalsize % Use small font for the table
    \begin{tabular}{|c|c|}
        \hline
        \textbf{\pauline{Parameter}} & \textbf{\pauline{Value}} \\ 
        \hline
        $R_{var}$ & 2.8 [$\Omega$] \\
        $L_{1,2}$ &  33 [$\mu H$] \\ 
        $C_{1,2}$ & 47 [$\mu F$] \\
        $R_{1,2}$ & 1 [$\Omega$]\\
        $V_{IN}$ & 40 [V] \\ 
        $C_{IN}$ & 120 [$\mu F$] \\ 
        $R_{C}$ & 0.1 [$\Omega$] \\ 
         $C_{out}$ & 240 [$\mu F$] \\ 
        $R_{IN}$ & 0.1 [$\Omega$] \\ 
       $R_{L_{1,2}}$ & 0.02 [$\Omega$]\\ 
        $R_{C_{1,2}}$ & 0.4 [$\Omega$] \\ 
        $R_{MOS_{on}}$&0.02 [$\Omega$]\\
        $\tau_{\text{meas}}$&0.2 [$\mu$ S] \\
        $\tau_{\text{comm}}$&2.5 [$\mu$ S] \\
        $T_{s}$&5 [$\mu$ S] \\
        $T_{PWM}$& 100 [$\mu$ S] \\ 
        ${\Delta t}_{PWM}$&0.1 [$\mu$ S] \\
        $T_{samp}$&100 [$\mu$ S] \\
        \hline
    \end{tabular}
    \label{tab:measurements}
\end{table}

% \begin{figure}[H] 
%     \centering
%     \includegraphics[width=0.5\textwidth]{Pics/timer_diagram.pdf}
%     \caption{\pauline{Detailed representation of the PWM generator.}}
%     \label{figure: Timer model}
% \end{figure}
%This type of converter efficiently steps down voltage levels, making it useful in various electronic devices like phone battery charger.

\begin{figure}[H]
    \centering
    \includegraphics[width=0.9\linewidth]{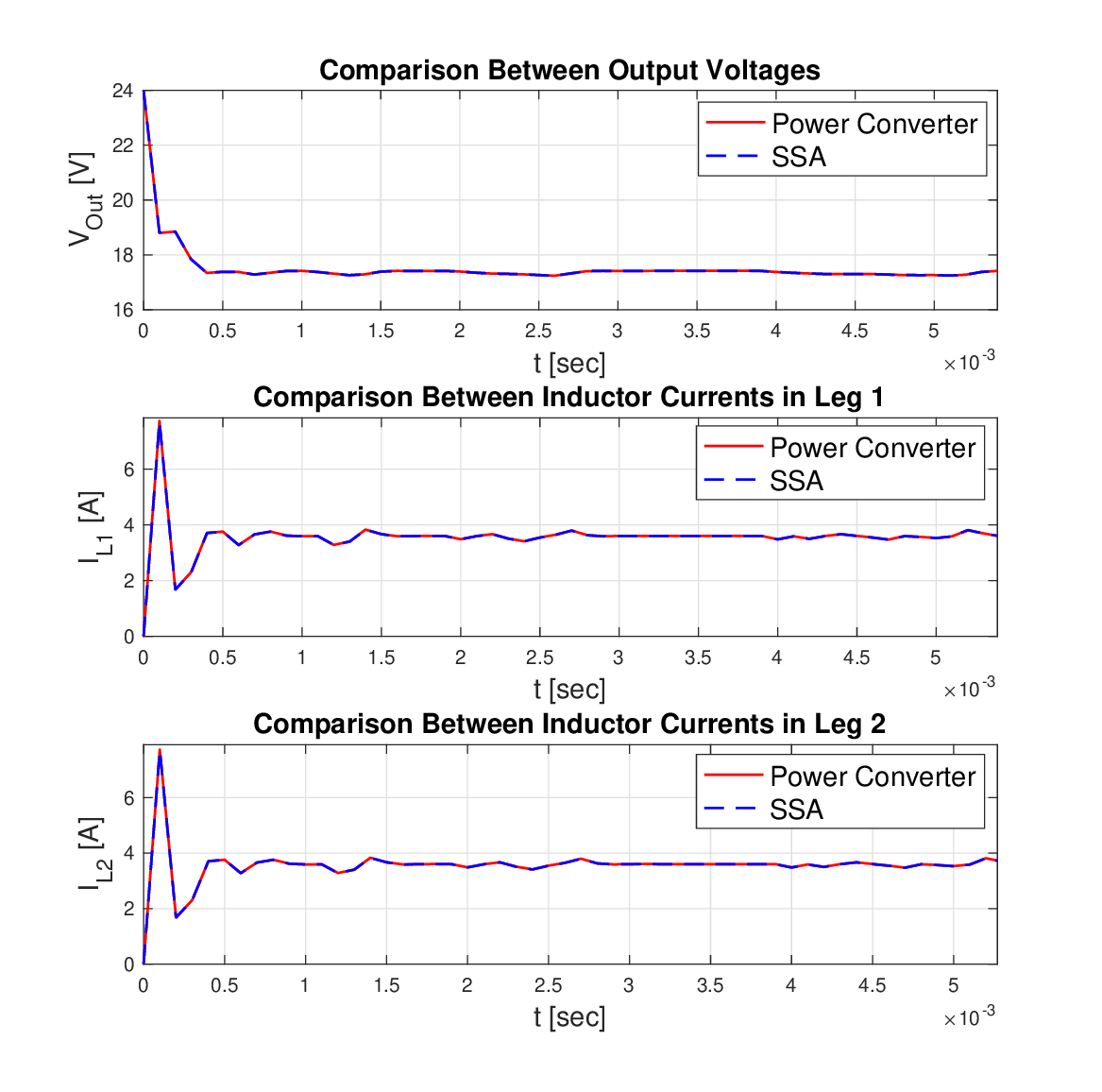}
    \caption{Comparison between SSA model and Power converter}
    \label{fig:SSA Comparison}
\end{figure}

%The switching operation found in the power converter results in changing the configuration of the components periodically, each configuration is described by separate set of equations, this leads to a difficulty in the transient analysis and the control design.
\pauline{Classically \cite{ref18}, an average state-space model can be obtained by applying Kirchhoff voltage law (KVL) and Kirchhoff current law (KCL) during the two operation modes. The state vector is $\bx=[\;I_{L_1} \;\; I_{L_2} \;\; V_{C_1} \; \; V_{C_2} \; \; V_{C_{IN}}\; \; V_{C_{out}}\;]^T$, the output is $y=V_{C_{out}}$, the controlled input is the duty-cycle $d(t)\in [0.1, 0.9]$ and the disturbance is $\bw=[V_{IN} \;\;\; \delta_{R_{var}}]^T$. The average state-space model can be expressed as :} 
\begin{equation}
\begin{array}{rl}
    \bK\dot{\bx}(t) &= \left( \bA_0 + d(t) \bA_1 \right) \bx(t) + \left( \bB_0 + d(t) \bB_1 \right) \bw(t) \\
    y(t) &= \bC \bx(t) 
\end{array}
\label{eq:2.26}
\end{equation}
\pauline{The matrices \( \bK, \ \bA_0, \ \bA_1 \in \mathbb{R}^{6 \times 6} \), \( \bB_0, \ \bB_1 \in \mathbb{R}^{6 \times 2} \), \( \bC \in \mathbb{R}^{1 \times 6} \), are given in Table \ref{table:matrices}. Figure \ref{fig:SSA Comparison} shows a comparison of the averaged state-space model (denoted as SSA) from \eqref{eq:2.26} and the Simulink circuit model from the OwnTech Foundation \cite{ref21} for an open-loop simulation with a constant duty-cycle $d=0.5$.} % \textcolor{red}{The figures below should all have the x-axis starting at 0!}
\pauline{In the present case, this model can be derived without too much difficulties as it remains a simple architecture, but this converter remains a good example of the complexity of the control task as the system is bilinear and subject to input saturation. This is usually handled by linearizing the model around a desired equilibrium point, the control design is then performed on the subsequent linear model. However, this does not allow to ensure that the performance and stability requirements are going to be met when the equilibrium point changes due to variations in the load or the input voltage. Lyapunov-based approach have shown promising results, see \cite{ahmad2025estimation} for instance, but they control directly the commutations of the switching devices, which make them not suited for PWM-based control, and their scalability is limited as the architecture becomes more complex and the state dimension increases. The next paragraph investigates the use of a data-driven control technique to overcome these challenges.}

% \textcolor{red}{Define K and give its dimension in the text + please in the future use labels that mean something for figures and equations, for instance the label eq2.26 could be replaced by eq:average-state-space-model}

% The matrix \( K \in \mathbb{R}^{6 \times 6} \) is a diagonal scaling matrix that accounts for component parameters and is defined as:

% where \( L_1, L_2 \) are inductances, and \( C_1, C_2, C_{IN}, C_{out} \) are capacitances associated with the system components.

\begin{table*}
\centering
\begin{minipage}{\textwidth}
    \scriptsize
    % \textcolor{red}{to be modified to fit the new equation format for simplification}
    $$A_0 =\left[\begin{array}{cccccc}
    -(R_{\text{on}} + R_{c_1} \parallel R_{1})& 0& -\frac{R_{1}}{R_{c_1} + R_{1}}& 0& 0& -\frac{R_{c_1}}{R_{c_1} + R_{1}} \\
    0& -(R_{\text{on}} + R_{2} \parallel R_{c_2})& 0& -\frac{R_{2}}{R_{c_2} + R_{2}}& 0& -\frac{R_{c_2}}{R_{c_2} + R_{2}} \\
    \frac{R_{1}}{R_{c_1} + R_{1}}& 0& -\frac{1}{R_{c_1} + R_{1}}& 0&0& \frac{1}{R_{c_1} + R_{1}} \\
    0& \frac{R_{2}}{R_{c_2} + R_{2}}& 0& -\frac{1}{R_{c_2} + R_{2}}& 0& \frac{1}{R_{c_2} + R_{2}} \\
    0& 0& 0& -\frac{1}{R_c + R_{\text{IN}}}& 0& 0 \\
    \frac{R_{c_1}}{R_{c_1} + R_{1}}& \frac{R_{c_2}}{R_{c_2} + R_{2}}& \frac{1}{R_{c_1} + R_{1}} &
    \hspace{4em} \frac{1}{R_{c_2} + R_{2}}& 0& -(\frac{1}{R_{c_1} + R_{1}} + \frac{1}{R_{c_2} + R_{2}} + \frac{1}{R_{var}})
    \end{array}\right]$$
    $$A_1 =
    \left[\begin{array}{cccccc}
    -R_c \parallel R_{\text{IN}} & -(R_c \parallel R_{\text{IN}})& 0 & 0 & \frac{R_{c_1}}{R_{c_1} + R_{1}}& 0 \\
    -(R_c \parallel R_{\text{IN}}) &  R_{\text{IN}} \parallel R_c  & 0 & 0 & \frac{R_{c_{2}}}{R_{c_2} + R_{2}} & 0 \\
    0 & 0 & 0 & 0 & 0 & 0 \\
    0 & 0 & 0 & 0 & 0 & 0 \\
    \frac{-R_{\text{IN}}}{R_c + R_{\text{IN}}} & \frac{-R_{\text{IN}}}{R_c + R_{\text{IN}}} & 0 & \frac{1}{R_c + R_{\text{IN}}} & -\frac{1}{R_c + R_{\text{IN}}}& 0 \\
    0 & 0 & 0 & 0 & 0 & 0
    \end{array}\right] \ \text{with } R_{on}=R_{L_i}+R_{MOS_{on}} $$

    $$K = 
    \left[\begin{array}{cccccc}
    L_1 & 0 & 0 & 0 & 0 & 0 \\
    0 & L_2 & 0 & 0 & 0 & 0 \\
    0 & 0 & C_1 & 0 & 0 & 0 \\
    0 & 0 & 0 & C_2 & 0 & 0 \\
    0 & 0 & 0 & 0 & C_{IN} & 0 \\
    0 & 0 & 0 & 0 & 0 & C_{out}
    \end{array}\right],
    \qquad
    B_0 =
    \left[\begin{array}{cc}
    \frac{R_c}{R_c+R_{\text{IN}}}  & 0 \\
    \frac{R_c}{R_c+R_{\text{IN}}} & 0 \\
    0 & 0 \\
    0 & 0 \\
    \frac{1}{R_c+R_{\text{IN}}}  & 0 \\
    0 & 1
    \end{array}\right]
    \qquad
    B_1 =
    \left[\begin{array}{cc}
    \frac{R_c}{R_c+R_{\text{IN}}}  & 0 \\
    \frac{R_c}{R_c+R_{\text{IN}}} & 0 \\
    0 & 0 \\
    0 & 0 \\
    0  & 0 \\
    0 & 0
    \end{array}\right]
    \qquad
    % B_0 =
    % \left[\begin{array}{cc}
    % 0 & 0 \\
    % 0 & 0 \\
    % 0 & 0 \\
    % 0 & 0 \\
    % \frac{1}{R_c+R_{\text{IN}}} & 0 \\
    % 0 & 1
    % \end{array}\right]
    % \qquad
    % B_1 =
    % \left[\begin{array}{cc}
    % \frac{R_c}{R_c+R_{\text{IN}}}  & 0 \\
    % \frac{R_c}{R_c+R_{\text{IN}}} & 0 \\
    % 0 & 0 \\
    % 0 & 0 \\
    % \frac{1}{R_c+R_{\text{IN}}}  & 0 \\
    % 0 & 1
    % \end{array}\right]
    % \qquad
    C =
    \left[\begin{array}{cccccc}
    0 & 0 & 0 & 0 & 0 & 1
    \end{array}\right]$$
    % \qquad
    % E_1 = E_0 =
    % \left[\begin{array}{cc}
    % 0 & 0
    % \end{array}\right]$
\medskip
\caption{Matrices of the average state-space model \eqref{eq:2.26}}
\label{table:matrices}
\hrule
\end{minipage}
\end{table*}

% \pauline{or equivalently as: }
%\begingroup
%\setlength{\abovedisplayskip}{2pt} % Space above equations
%\setlength{\belowdisplayskip}{2pt} % Space below equations
% \begin{align}
%     K\frac{dx(t)}{dt} &= A'x(t) + N(x \otimes d(t)) + B(W \otimes d(t)) + B_w W + \notag \\
%     &\quad N_w(x \otimes W) \\
%     y(t) &= C'x(t) + M(x \otimes W) 
% \end{align}
% %\endgroup
% \textcolor{red}{Isn't it A,  and C instead of A' and C'?}

% \pauline{where $\otimes$ represents the Kronecker product : }
% \begin{align}  
% &x\otimes d=[I_{L_1}d\; \; I_{L_2}d\; \dots V_{C_{out}}d]^T  \\
% &x\otimes u=[I_{L_1}V_{IN} \; \dots V_{C_{out}}V_{IN},\;I_{L_1}\delta_R\; \dots V_{C_{out}}\delta_R]^T  \\ 
% &W\otimes d=[V_{IN} d\; \  \delta_R d]^T 
% \end{align}

\section{Data-Driven Control of power converters}
\pauline{The difficulty of traditional model-based control increases as power converters grow more complex.} %, traditional model-based control methods become increasingly impractical due to the difficulty of developing accurate mathematical models for such systems. 
Data-driven control methods offer a promising alternative by designing controllers directly from data without relying on explicit models. In particular, among the various data-driven approaches, \emph{Ziegler-Nichols} and \emph{Virtual Reference Feedback Tuning} (VRFT) \pauline{are popular techniques that are going to be used in this section.} %stand out as effective techniques for systems like power converters.\\ 
\pauline{The controller $K$ to be designed is a PI of the following form:}
\begin{equation}
    K(s) = K_p + K_I\frac{1}{s}
    \label{eq:PIstructure}
\end{equation}
\pauline{where $K_p$ is the proportional gain and $K_I$ is the integral gain.}

% \textcolor{red}{Can you check that this is the structure you used in VRFT? Looking at step 3 of your algorithm it seems you enforce a PD structure instead of a PI.}

\subsection{\pauline{Baseline control using Ziegler-Nichols (ZN) tuning}}
\pauline{The \emph{Ziegler-Nichols} (ZN) method is a heuristic that traces back to the 1940s \cite{ziegler1942optimum}. The proportional gain is progressively increased until the system exhibits stable and sustained oscillations : the ultimate gain $K_u$ and the oscillations period $T_u$ are then measured. Several simulations have been conducted on the Owntech Simulink model to get $K_u=0.065$ and $T_u=1\;[ms]$. The controller gains can then be fixed using the following tuning rule :}
\begin{equation}
    K_p= 0.45 K_u\text{ and } K_I =\frac{0.54K_u}{Tu} 
    \label{eq:ZN_tuning_rule}
\end{equation}

% \textcolor{red}{You obtained a continuous controller that you discretized, is that it? which sampling time and method did you use? Also where did you get that tuning rule ? I have never seen $K_p=K_u$ before} 

%The integral gain 
%$K_I$ is then determined using the formula for optimal performance: $K_I=\frac{1.2K_P}{T_u}$, where $T_u$ is the oscillation period. After conducting several trials on the OwnTech's power converter, the optimal values for the gains were determined: the proportional gain $K_p=0.0082$, and the integral gain $K_i=66.61$. 

\pauline{The ZN tuning rules were initially proposed for the design of a continuous controller for a system that can be approximated by a transfer function of the following type:}
\begin{equation}
    k\frac{e^{-\tau s}}{s+a}
    \label{eq:TF_nyquist}
\end{equation}
\pauline{However, not all systems can be correctly approximated by such transfer function, which constitute the main limitation of the ZN tuning rules and the gains are tuned in a rather aggressive way, which can cause problem when discretizing the controller.}

\subsection{\pauline{Virtual Reference Feedback Tuning (VRFT)}}
\begin{figure}[H] 
 	\includegraphics[width=0.9\linewidth]{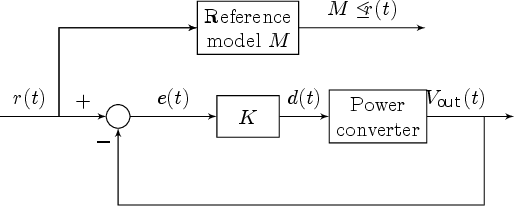}
    \caption{Principle of VRFT\cite{ref5}}
    \label{fig:VRFT Princ}
\end{figure}

\pauline{The VRFT was initially proposed in \cite{ref5} and with (up to our knowledge) two successful application to power converters in \cite{remes2020virtual} (without anti-windup) and \cite{ref34} (on a simpler model). The control design is formulated as a controller identification problem by using a reference model to enforce the specifications. This method is entirely data-driven, and only measurements data from a single experiment are needed to get a controller.} This process simplifies the controller design process %by using a virtual reference signal to identify the behavior of the desired closed-loop system. This eliminates the requirement for a 
as a precise yet simple enough mathematical model of the system is no longer needed.
%In this section, VRFT is applied to the power converter provided by the OwnTech Foundation \cite{ref21}. The VRFT method is employed to design a controller directly from experimental data, without the need of an explicit mathematical model of the converter.  \\
%The Virtual Reference Feedback Tuning (VRFT) method is a data-driven approach. 
The principle of VRFT is represented on Figure \ref{fig:VRFT Princ}. The objective of VRFT is to design a controller $K$ that closely approximates the ideal controller $K^\star$. This ideal controller is the one ensuring that the closed-loop transfer function matches the desired reference model $M$. %the  when both the reference model $M$ and the closed-loop system with the controller are fed with the virtual reference signal $r(t)$.
\pauline{The VRFT follows the five steps detailed hereafter :}
\begin{itemize}
    \item \textbf{Step 1: Data collection} 
    \item[] An experiment is conducted on the power converter, with a user-defined input signal (here the duty cycle signal $d(t)$), and the resulting output voltage $V_{out}(t)$ is collected. \pauline{In order for the control design to be efficient, the chosen input signal should excites the dynamics to be controlled. In this paragraph, the training data for the VRFT is obtained by an open-loop simulation of the Owntech Simulink model with a chirp duty cycle varying around 0.5 with a frequency varying from $f_c/2$ to $2f_c$. The collected output is artificially corrupted by a noise signal $\eta(t)$, with $|\eta(t)|\le 0.5$ . The collected training data is visible on Figures \ref{fig:training input data} and \ref{fig:training output data}.} The dataset used for training consists of $N=501$  samples collected over a simulation duration of 0.05 seconds. 
    %\textcolor{red}{This level of noise is meaningless, see the scale of Vout on the figure. You should increase it or we remove it from the paper. What's the length $N$ of the dataset and the duration of the training experiment?}
    \begin{figure}[h]
        \centering
        \includegraphics[width=0.85\linewidth]{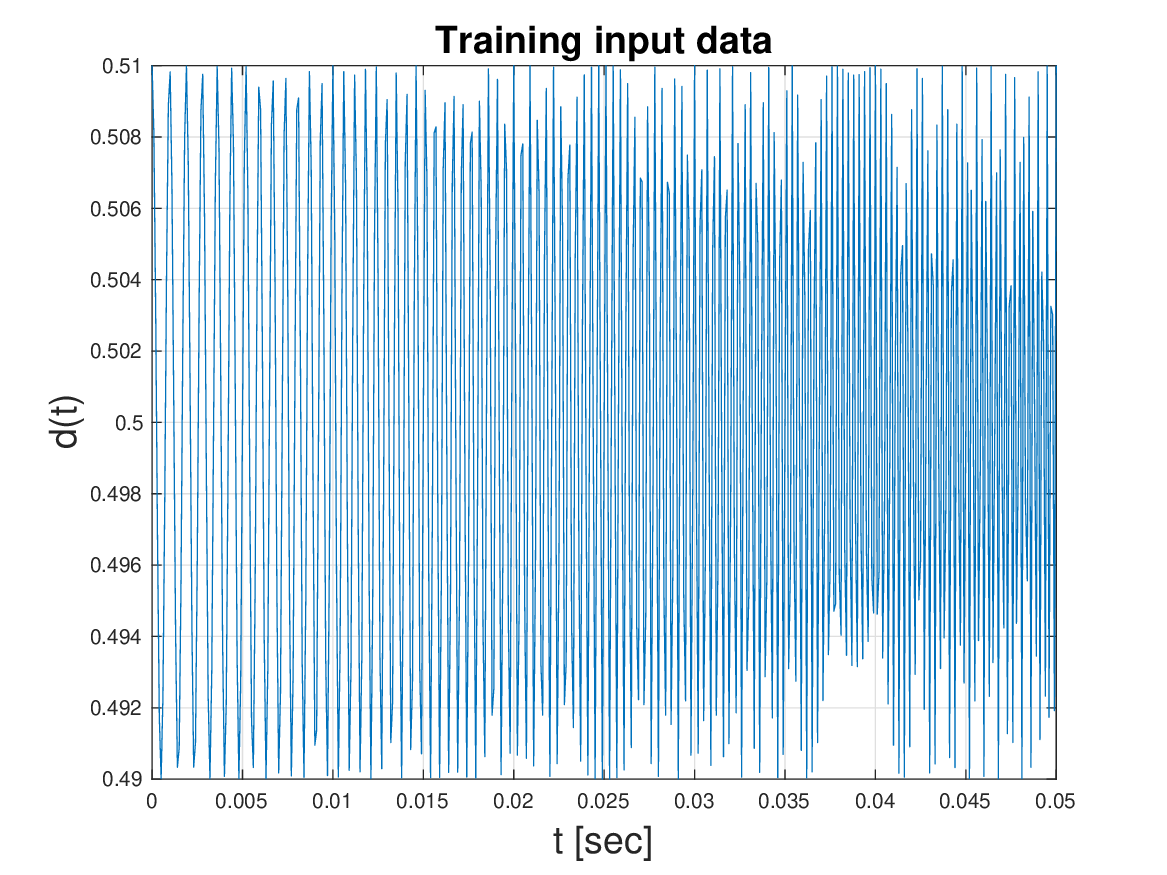}
        \caption{Input $d(t)$ used for data collection.}
        \label{fig:training input data}
    \end{figure}
    
    \begin{figure}[h]
        \centering
    \includegraphics[width=0.85\linewidth]{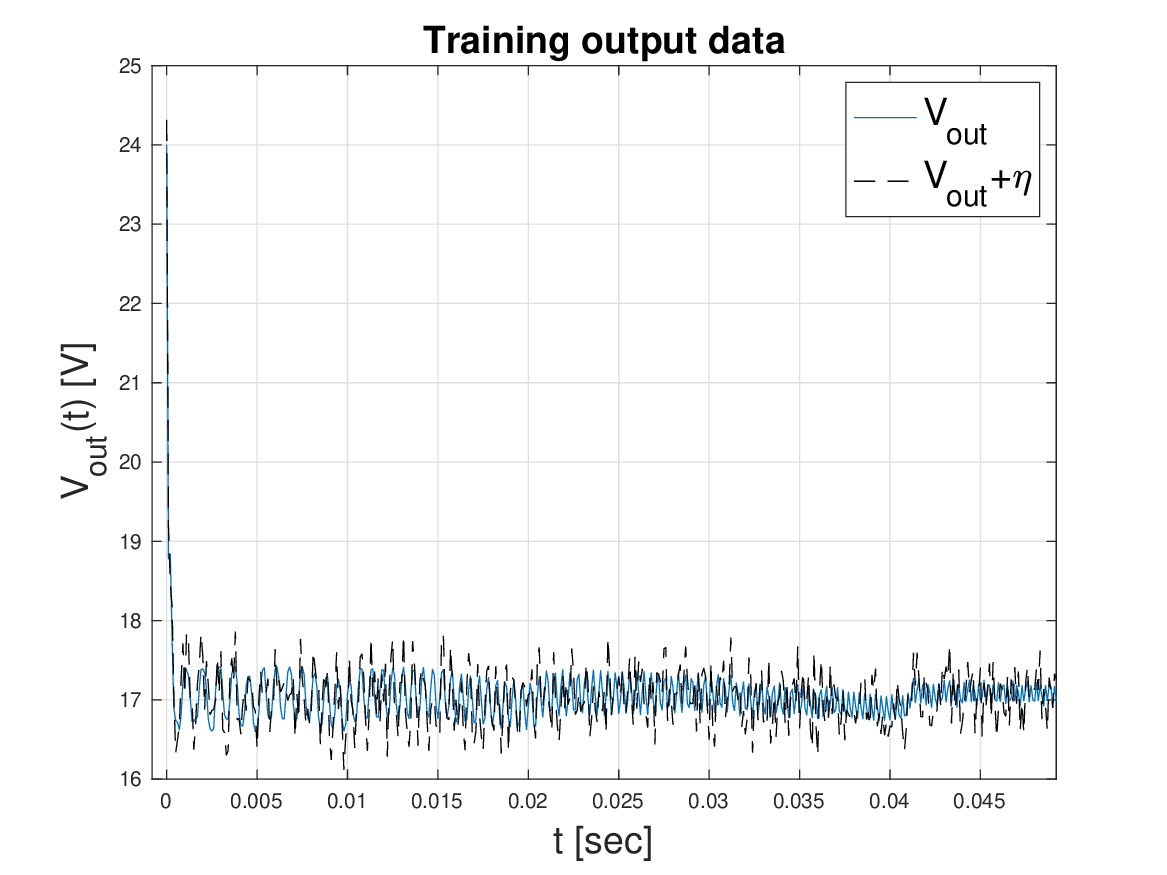}
        \caption{Collected output $V_{out}(t)$. }
        \label{fig:training output data}
    \end{figure}
    \item \textbf{Step 2: Specifications}
    \item[] The desired reference model $M$ is defined to represent the desired closed-loop behavior. \pauline{In the present case, a simple first-order model, see \eqref{eq:ref_model} is used to enforce a time constant  $\tau=\frac{1}{f_c}$ corresponding to a desired cut-off frequency $f_c$ (here we take $f_c=f_s/100$ with $f_s=1/T_s=200kHz$ the switching frequency):}
    \begin{equation}
        M(s)=\frac{1}{1+s\tau}
        \label{eq:ref_model}
    \end{equation}
    The reference model is discretized using the ZOH method with a sampling time $T_{samp}=100 \;[\mu \;s]$.
    \pauline{It should be noted that the choice of a reference model can be tricky when the system to be controlled is unstable or non-minimum phase, see \cite{kergus2019from}.}
    \item\textbf{Step 3: Control structure} 
    \item[] \pauline{The controller to be tuned is expressed as $K(z,\theta)=\beta^T(z)\theta$, where $\beta(z)$ defines the structure of the controller and is defined in discrete time. The parameters to be optimized are grouped in the vector $\theta$. In the present case, the controller is a PI, which gives} $\theta= [K_p\;\; K_I]^T$ and
    \begin{equation}
        \beta^T=\begin{bmatrix}
            1& \frac{z}{z-1}
        \end{bmatrix}
    \end{equation}
    \pauline{It should be noted that, in theory, the control structure $\beta(z)$ should be chosen so that it can approximates correctly the ideal controller $K^\star$, which is a complicated task when the system to be controlled is unknown and/or complex.}
    \item \textbf{Step 4: Control tuning} 
    \item[] \pauline{First, the virtual reference signal $r(t)$ is constructed so to satisfy $V_{out}(t)=M(z)\cdot r(t)$. It corresponds to the reference that would have been sent to the desired closed-loop model in order to get the training output $V_{out}(t)$. The corresponding virtual tracking error $e(t)=r(t)-V_{out}(t)$ is then computed. The ideal controller $K^\star$ is then the one that gives the training signal $d(t)$ when it is fed with the virtual error $e(t)$ : this is how the control tuning problem is recast as controller identification.} %(Assuming $M(q^{-1})\neq 1$, else the virtual tracking  error is zero)
    \begin{figure}[H]
        \centering
        \includegraphics[width=0.85\linewidth]{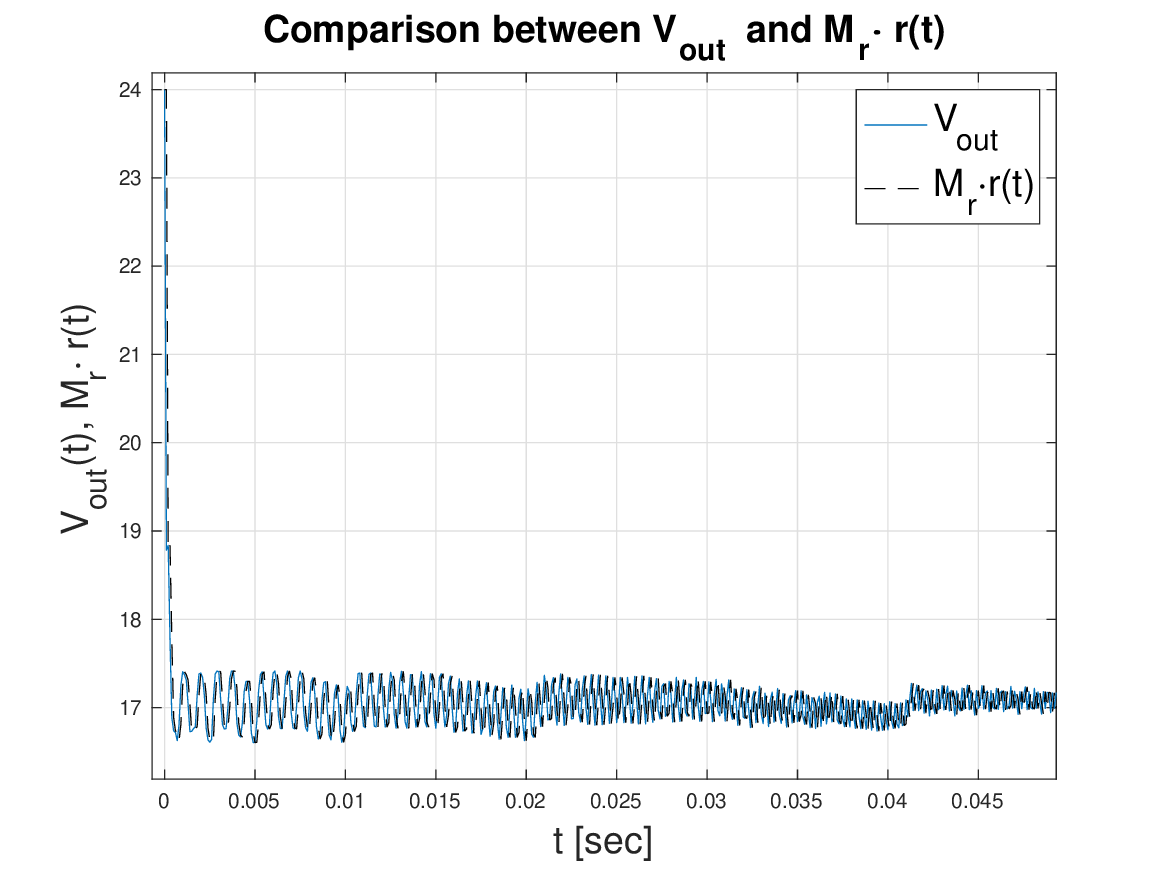}
        \caption{Comparison between $V_{out}(t)$ and $M\cdot r(t)$}
        \label{fig:ref and output comparison}
    \end{figure}
    The controller parameters $\theta = [K_p\;\;K_I]^T$ are then found by minimizing the following criterion
     \begin{align}
       J_{VRFT}=\frac{1}{N} \sum_{t=1}^{N}(d(t)-K(z;\theta)\cdot e(t))^2 \\ \notag
       =\frac{1}{N} \sum_{t=1}^{N}(d(t)-\Phi^T (t)\cdot \theta)^2 
    \end{align} 
     Knowing that $K(z,\theta,)=\beta^T(z)\theta$, the regressor is $\Phi(t)=\beta(z)e(t)$. This is a least squares problem that can be solved easily. The solution $\theta^\star$ is given by: 
      \begin{equation}
          \theta^\star=[\sum_{t=1}^{N}\Phi(t)\cdot \Phi(t)]^{-1}\sum_{t=1}^{N}\Phi(t)\cdot d(t)
      \end{equation}
    In the end, the following parameters are estimated :
           \[ \theta^\star=\left[ K_p^\star \ K_I^\star \right]^T=\left[ 0.0031 \; \;0.0065\right]^T    \]
    After discretization, the  PI controller implemented in the Simulink model is the following : 
    \begin{align}
        K(s) = 0.0031+ 0.0065\cdot
        \frac{T_s}{s}
    \end{align}
    % \begin{align}
    %         K(z, \theta^\star) = 0.0041 + 
    %         \frac{0.0064\cdot q^{-1}}{1-q^{-1}}
    %     \end{align}
\end{itemize}

\begin{figure}[H]
    \centering
    \includegraphics[width=0.9\linewidth]{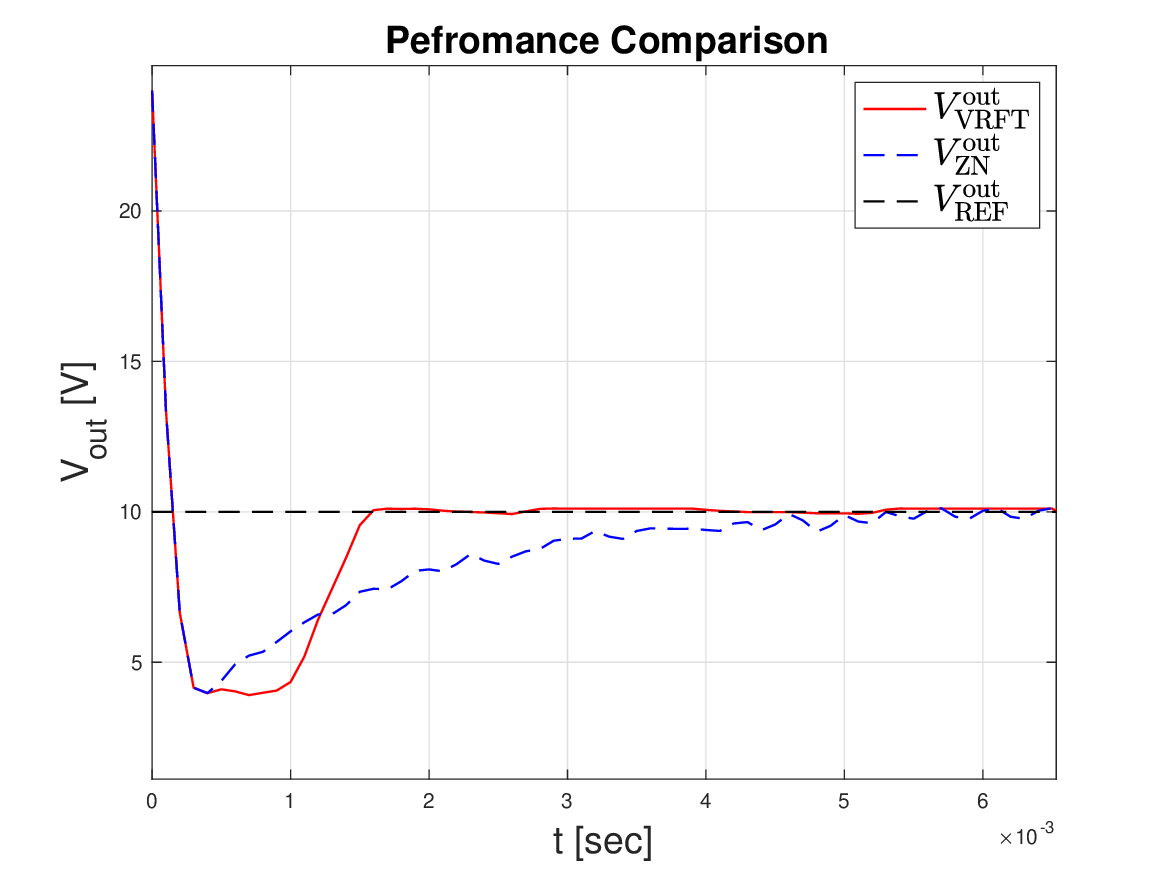}
    \caption{Closed-loop performance comparison between the ZN and VRFT controllers with a constant reference signal }
    \label{fig:Perfomrance comparison}
\end{figure}

\begin{figure}[H]
    \centering
    \includegraphics[width=0.9\linewidth]{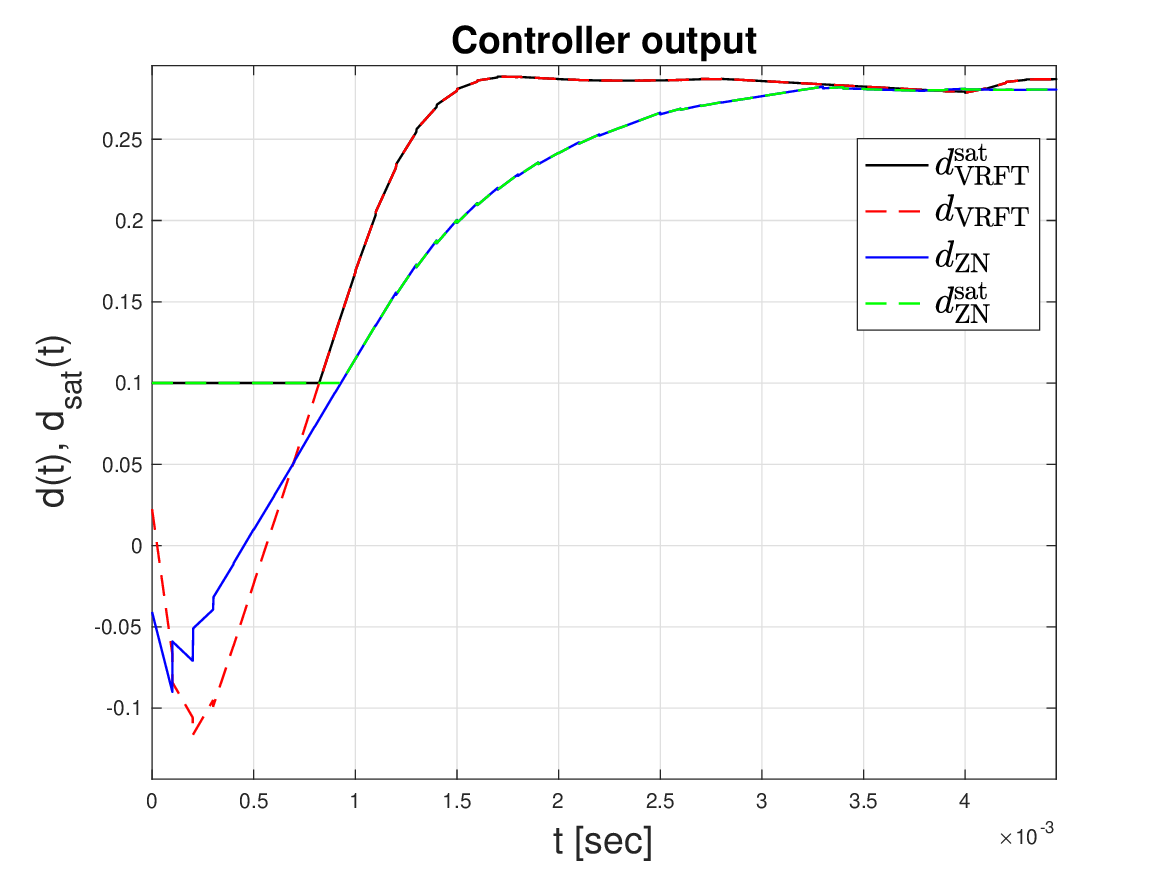}
    \caption{Comparison between the ZN and VRFT controllers' output  with the saturated duty cycle. }
    \label{fig:Duty comparison}
\end{figure}

Figure \ref{fig:Perfomrance comparison} show the performances of the ZN and VRFT controllers for a constant $V_{out}^{ref}=10 [V] $ \pauline{and the corresponding control signals are represented on Figure \ref{fig:Duty comparison}. While one can see that the VRFT controller performs better than the ZN one with a faster response, a significant undershoot is observed for both controllers.}

The duty cycle of the OwnTech converter is constrained within the range \( [0.1, 0.9] \) (see Figure \ref{fig:control_structure}), which explains the large undershoot observed in Figure \ref{fig:Perfomrance comparison} : this is due to the saturation of the duty cycle observed in Figure \ref{fig:Duty comparison}. This phenomenon is known as \textit{windup}, and it is commonly encountered with PI controllers. When the controller output exceeds the duty cycle saturation limit, the integral term continues to accumulate the error, resulting in an integrated error buildup. As noted in \cite{ref33}, this issue becomes critical as the accumulated error can lead to significant undershoot and, in some cases, may cause unstable behavior. \pauline{In the present case, the ZN controller leads to a longer saturation than the VRFT one. }

%\textcolor{red}{Please remove the $(t)$ in the legend and on Figure \ref{fig:Perfomrance comparison} and \ref{fig:AW Perfomrance} it should be $V_{ou}^{ref}$ instead of $r$ in the legend}
\vspace{-0.4cm}

\subsection{Extended VRFT With Anti-Windup Compensation} 
\begin{figure}[H]
    \centering
    \includegraphics[width=0.9\linewidth]{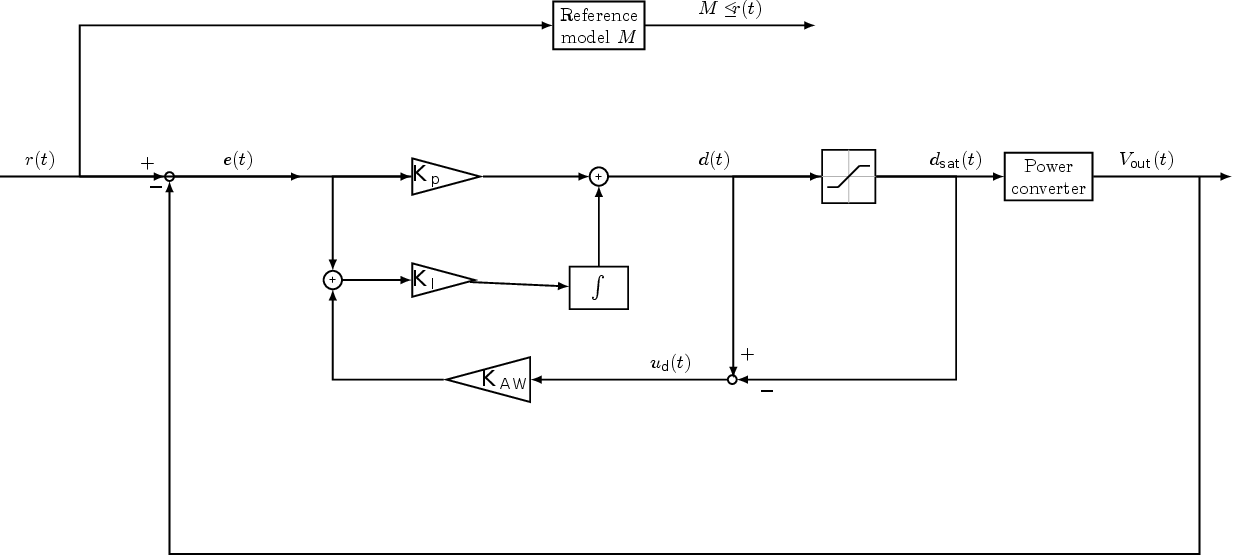}
    \caption{The PI controller with Anti-Windup}
    \label{fig:AW princ}
\end{figure}

% This section describes how to
% extend VRFT toward the digital PI controller
% that includes linearly approximated anti-windup compensation. \\
\pauline{In order to improve the closed-loop performances, an anti-windup compensation is included in the designed controller, following the VRFT extension proposed in \cite{breschi2020direct}.}
Anti-windup is commonly used  in control systems to prevent the integrator from accumulating error when the controller output becomes saturated. While there are several anti-windup methods, the one adopted here \cite{ref20} uses the difference between the saturated
input and the actual control input to regulate the behavior of the integral action in order to avoid windup phenomenon. \pauline{Such control structure is illustrated on Figure \ref{fig:AW princ}. The duty cycle $d_{sat}(t)$ that is actually sent to the power converter is given by : }
\begin{equation}
    d_{sat}(t)=\begin{cases}
    0.9 & \text{if } d(t) > 0.9 \\
    0.1 & \text{if } d(t) < 0.1 \\
    d(t)& \text{otherwise}
    \end{cases}
\end{equation}

% Compensating the anti-windup in the structure of the controller will lead to change of the optimization problem in Equation (\ref{eq:opt}): 
% \begin{equation}
%     J_{VR}(\theta)=||u(t)- \Phi_{f}(t) \theta_f||^2
%     \label{eq:opt2}
% \end{equation}
% \textcolor{red}{You need to add $e_{int}$ to your figure to define this notation. What's $u_d$?}
As shown in Figure \ref{fig:AW princ}, the controlled output $d(t)$ is given by:
\begin{equation}
    d(t)=(K_p \cdot e(t))+K_I\int e(t)dt+K_IK_{AW}\sum_{j=1}^{n_{Aw}}u_d(t-j)
    \label{eq:CK}
\end{equation}
% Where $e_{int}(t)$
% represents the integral of the control error over time. It addresses the steady-state error accumulated by the error signal $e(t)$ over time. The integral of the error is updated recursively at each time step $t$:
% \begin{equation}
%     e_{int}(t)=e_{int}(t-1)+e(t)
% \end{equation}
% The integral term provides memory to the controller, capturing past errors and driving the control input $u(t)$. By adding this accumulated error term, the system is pushed towards zero steady-state error.
where $u_d=d-d_{sat}$ represents the anti-windup term. \pauline{It allows to act on the control signal only when the PWM generator saturates, in which case $u_s\neq 0$ while $u_d=0$ otherwise. The design parameter $n_{AW}$ determines how many past values of $u_d$ are taken into account in the anti-windup compensation.} %number of anti-windup terms $n_{AW}$ is a design parameter. It determines how many past values of the difference between the  control signal $u(t)$ and its saturated version $u_r(t)$ are considered in the controller. 
As in \cite{ref20}, $n_{AW}=1$ is selected, which is often sufficient. A larger $n_{AW}$can be selected for slow systems where saturation effects persist over multiple time steps, but it increases computational complexity and the number of parameters to tune.
% (If the system has long transients or slow dynamics, a higher $n_{AW}$ is considered. Larger $n_{AW}$ helps the system ) \\

\begin{figure}[h]
    \centering
    \includegraphics[width=0.85\linewidth]{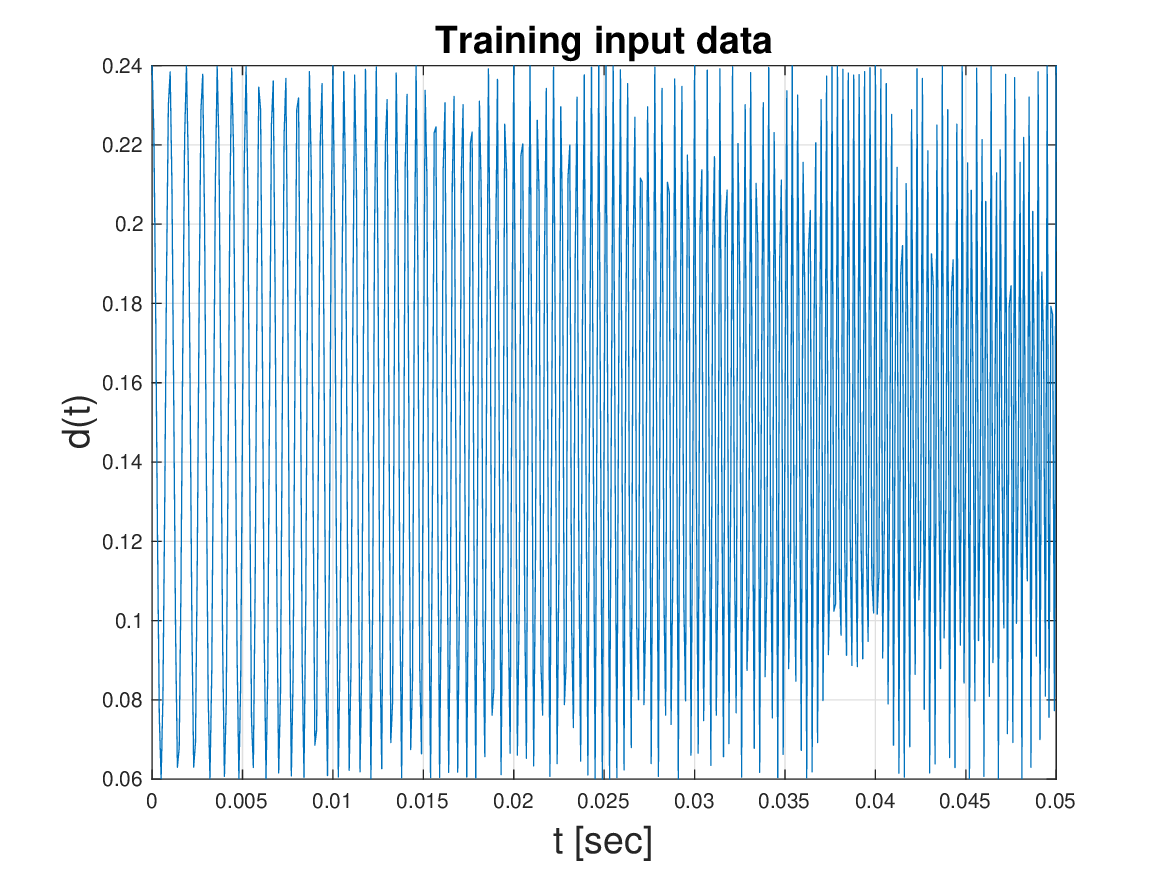}
    \caption{Input $d(t)$ used for data collection for the VRFT design including anti-windup compensation.}
    \label{fig:d(t)}
\end{figure}

\begin{figure}[h]
    \centering
    \includegraphics[width=0.85\linewidth]{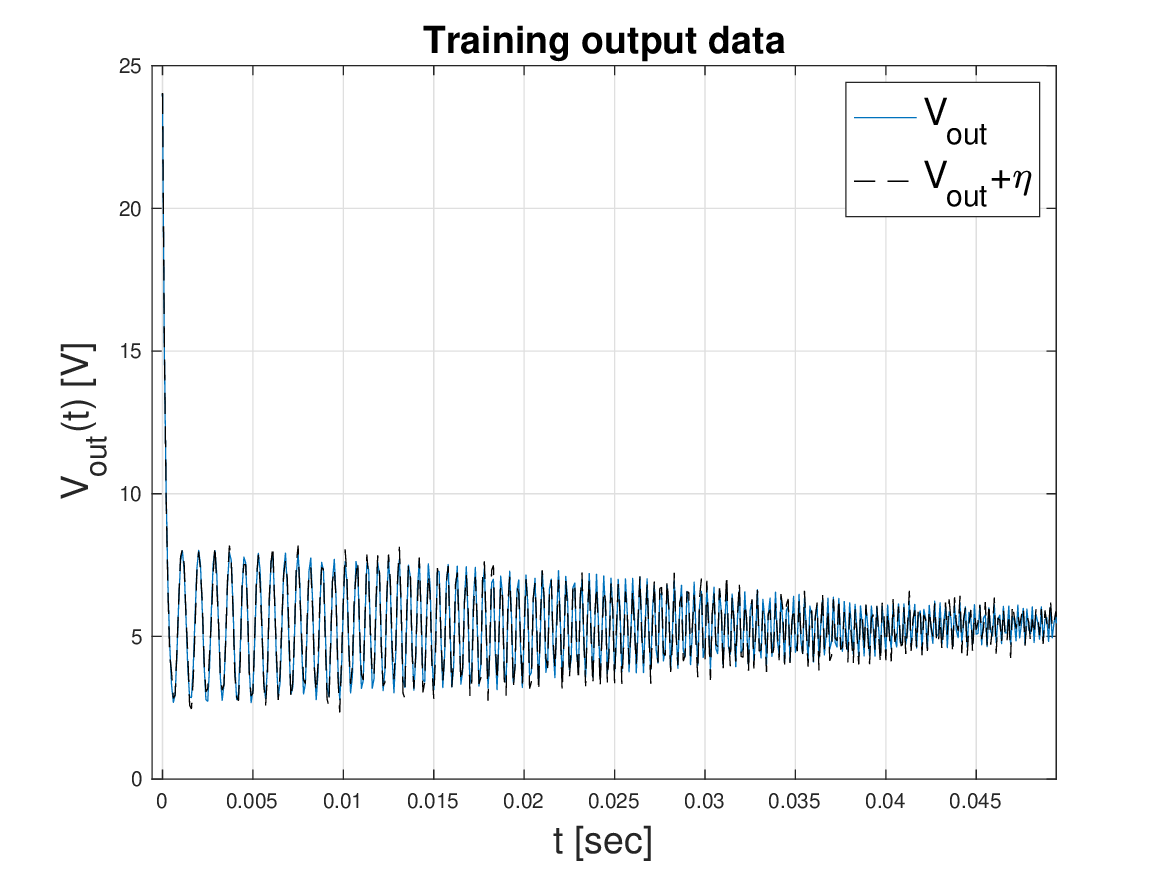}
    \caption{Collected output $V_{out}(t)$ for the VRFT design including anti-windup compensation.}
    \label{fig:y(t)}
\end{figure}

\begin{figure}[h]
    \centering
    \includegraphics[width=0.85\linewidth]{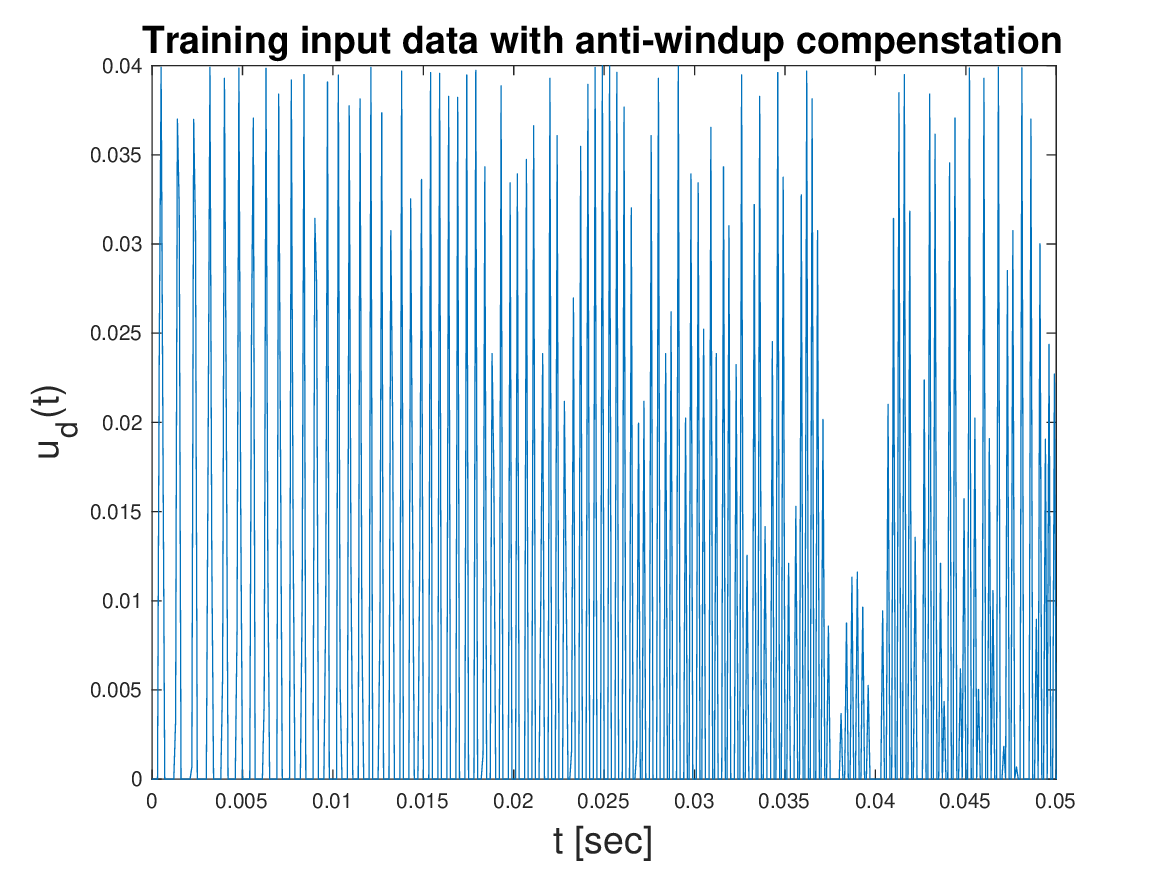}
    \caption{Collected new controller input $u_d(t)$ for the VRFT design including anti-windup compensation.}
    \label{fig:ud(t)}
\end{figure}

\pauline{Compared to the classical VRFT approach recalled in the previous paragraph, the collected data includes now the signal $u_d$ in addition to the input-output measurements. Note that to design the ant-windup compensation, it is necessary that the training experiment hits saturation. For this reason, compared to the previous paragraph, the training data set is modified and is visible on Figures \ref{fig:d(t)}, \ref{fig:y(t)} and \ref{fig:ud(t)}. The data collection simulation is perfomed in open-loop, with a chirp input signal $d(t)$ spanning the same frequency range as before, but around a fixed operating point of $0.15$ to ensure that the duty cycle goes below the saturation limit $0.1$, leading to the occurrence of the windup phenomenon which is part of the dynamics to be controlled (see the traing set for $u_d$ on Figure \ref{fig:ud(t)}.}
%\textcolor{red}{Can you plot the training $u_d$ you use in the VRFT-AW design? It's part of the trainign data in that case}

\pauline{The controller structure $\beta(z)$ is also changed to include the anti-windup compensation from Figure \ref{fig:AW princ}.}
\pauline{The virtual reference $r$ and the virtual tracking error $e$ are then computed as previously. The problem can then be formulated as a least-squares problem as in \textbf{Step 4} of the classical VRFT, but with new regressors :}% shown in Equation~(\ref{eq:v_error}). Similarly, the virtual reference signal is calculated as described earlier in Equation~(\ref{eq:v_ref}).
 \begin{equation}
    J_{VRFT,AW}(\theta_{AW})=||d- \Phi_{AW} \theta_{AW}||^2
    \label{eq:opt2}
\end{equation}
with the regression matrix $\Phi_{AW}$ is :
\begin{equation}
\Phi_{AW}(t) =
\begin{bmatrix}
e_(t) & \int e(t)dt & u_d(t-1) 
\end{bmatrix}^\top
\end{equation}
and the parameters to be estimated $\theta_{AW}$ :
    \begin{equation}
    \theta_{AW}=[K_p \;\;\; K_i\;\;\; K_{AW} K_i]^T
  \end{equation}
The optimal parameter vector $\theta_{AW}$ that minimizes \eqref{eq:opt2} is then given by: 
\begin{equation}
  \theta_{AW}^\star=[\sum_{t=1}^{N}\Phi_{AW}(t) \Phi_{AW}(t)]^{-1}\sum_{t=1}^{N}\Phi_{AW}(t) d(t).
\end{equation}

% For the training output data, the input signal $d(t)$ is applied to the power converter, and the corresponding output data $V_{out}(t)$ is collected.

The estimated controller has the following gains :% \textcolor{red}{to be completed}
\begin{equation}
    K_{p}^\star = 0.0018, \ K_I^\star= 0.0056 \text{ and } K_{AW}^\star=441.47
\end{equation}
% \begin{equation}
%     K_{AW}(z, \theta) = 0.0015 + \frac{ 0.0051\cdot q^{-1}}{1-q^{-1}},\;\;\;K_{AW}=439.47
% \end{equation}   
% \subsection{Simulation Results}
 
\begin{figure}[h]
    \centering
\includegraphics[width=0.9\linewidth]{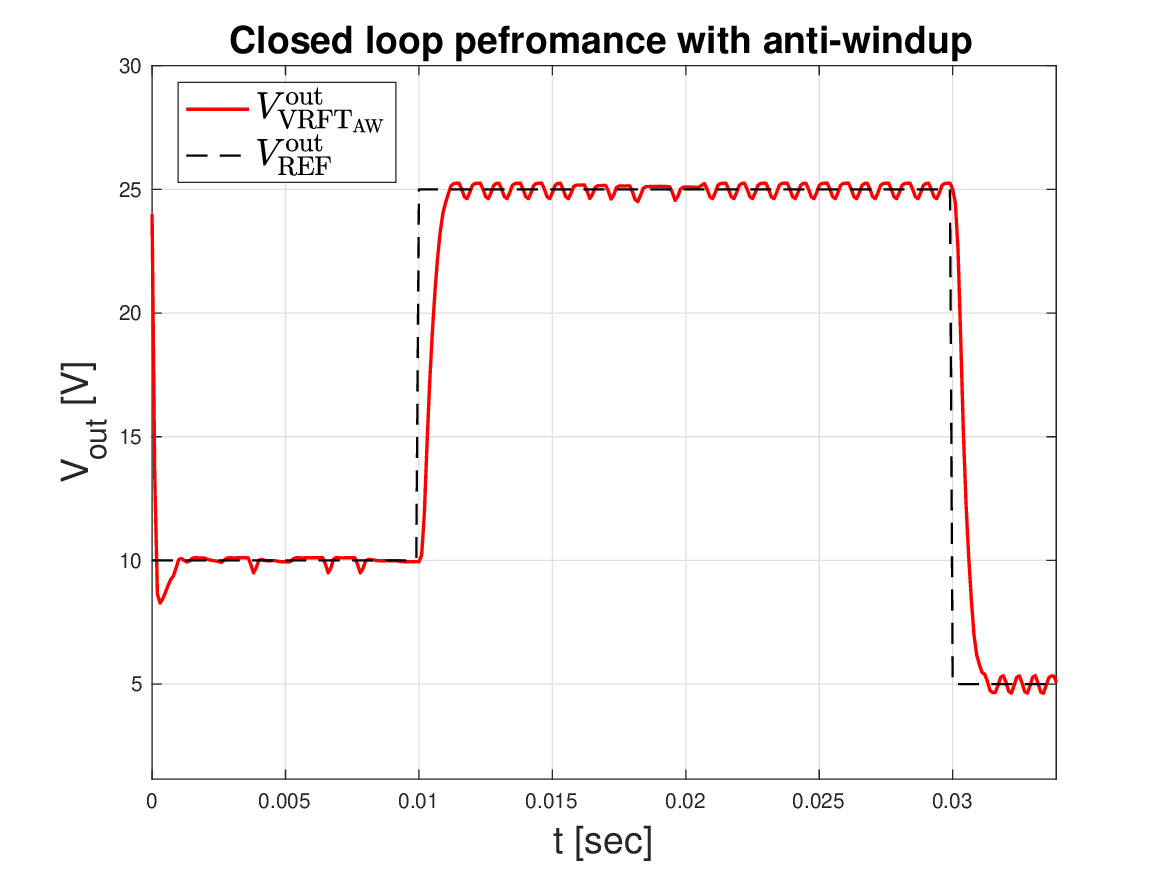}
    \caption{Closed-loop performance with anti-windup implementation }
    \label{fig:AW Perfomrance}
\end{figure}

%you need a figure that compare the VRFT without anti-windup and with anti-windup, and ZN. Please simulate the three closed loops for the same reference signal and \textcolor{red}{ do not forget to plot also the control signal (non saturated) and the saturated one, for all cases also.}
\begin{figure}[h]
    \centering
\includegraphics[width=0.9\linewidth]{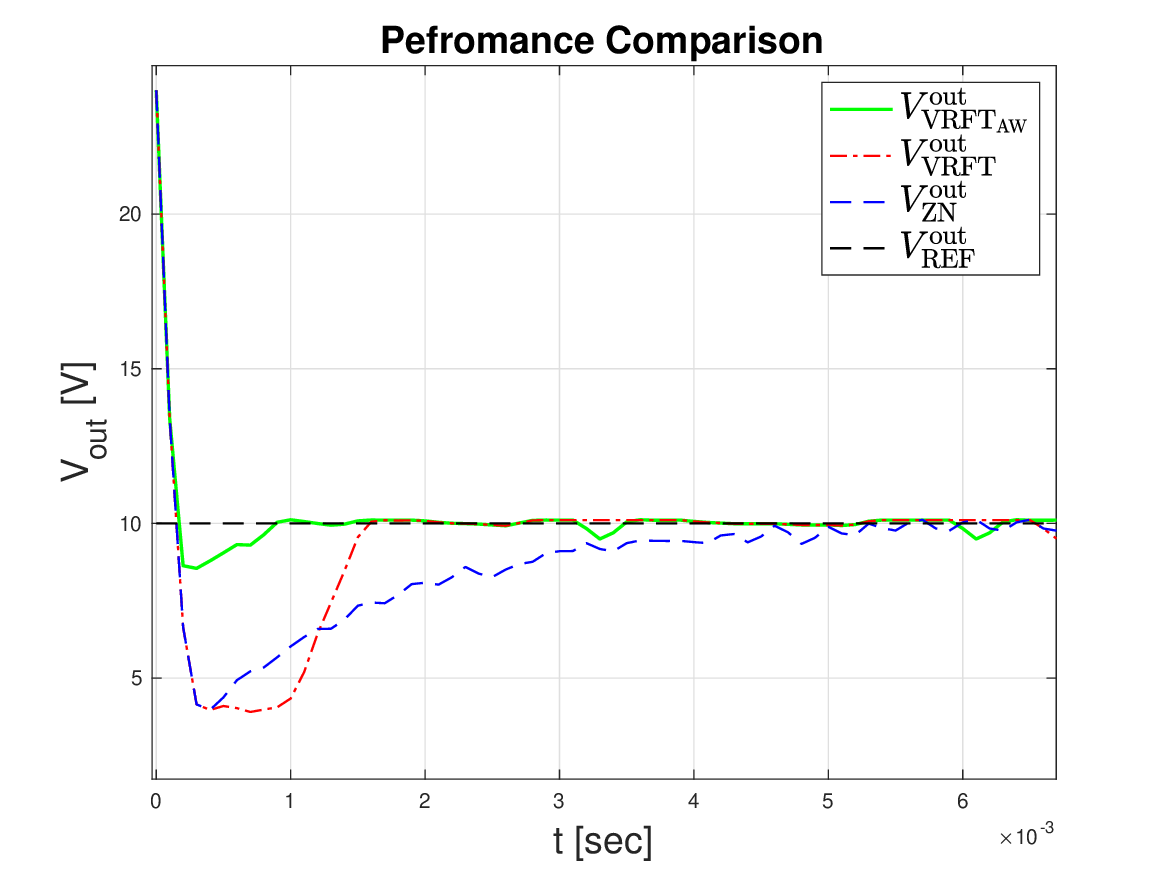}
    \caption{Closed-loop performance comparison}
    \label{fig:Final Perfomrance}
\end{figure}
\begin{figure}[h]
    \centering
    \includegraphics[width=0.9\linewidth]{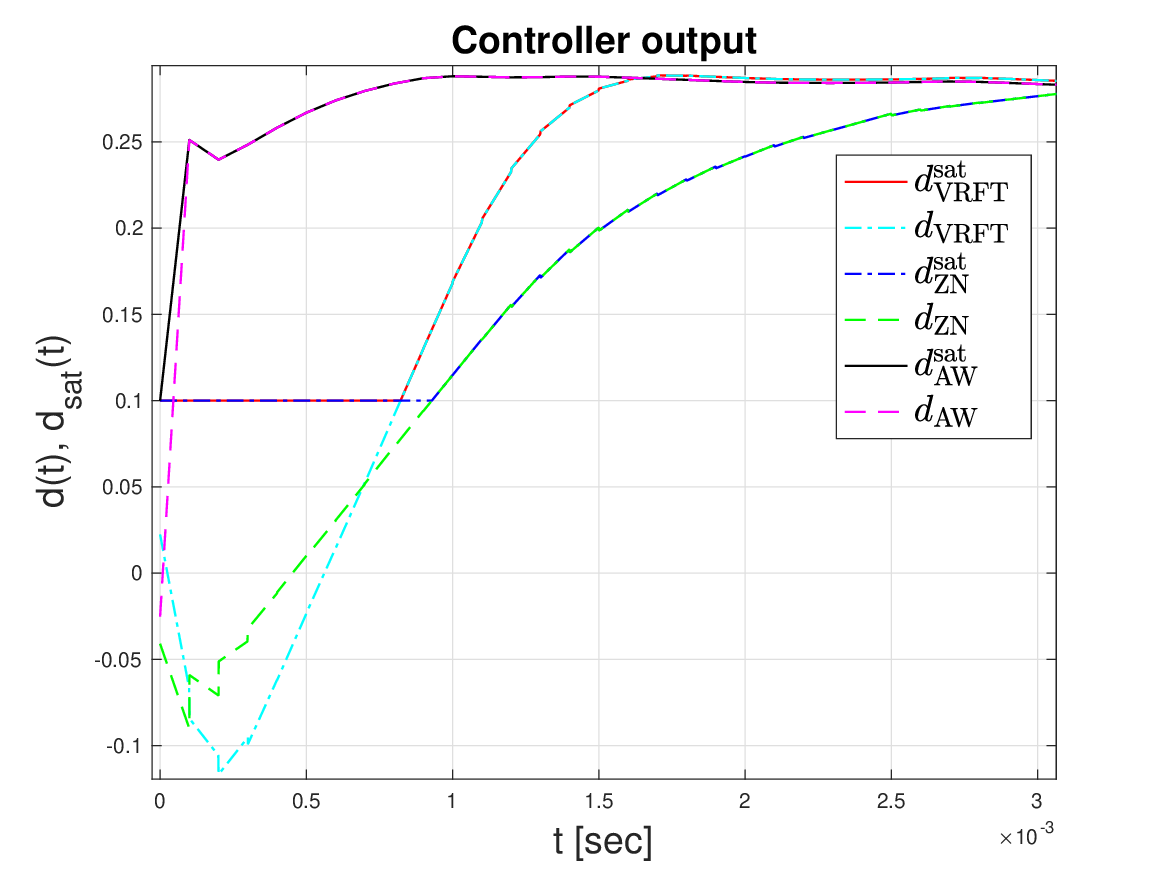}
    \caption{Comparison between the controllers' output  with the saturated duty cycle. }
    \label{fig:Duty2 comparison}
\end{figure}
As visible on Figures \ref{fig:Final Perfomrance} and Figure \ref{fig:Duty2 comparison}, the \pauline{tracking} error is reduced by \pauline{including an anti-windup compensation, which allows to adjust the integral gain. Performances of the different designed controllers are summarized in Table \ref{tab:perf_indicators}, showing that the VRFT data-driven approaches used in this paper outperform the baseline controller obtained by the ZN tuning rules (the performance indicators are calculated on the first transient to reach $V_{out}^{ref}=10V$, see Figure \ref{fig:Final Perfomrance}).}

%\textcolor{red}{In the table, you calculated the indicators only for the first transient to reach $V_{out}=10V$, is that correct?}

% \begin{table}[h]
% \scriptsize
%     \centering
%     \caption{Closed Loop Performance Comparison}
%     \begin{tabular}{|c|c|c|c|}
%         \hline
%         Controller & Undershoot (\%) & Rise Time (ms) & Settling Time $5$ (ms) \\
%         \hline
%         $C_{ZN}$ & 61 & 1.4 & 2.7 \\
%         $C_{VRFT}$ & 60.8 & 0.84 & 1.8 \\
%         $C_{VRFT_{AW}}$ & 11.4 & 0.8 & 0.9 \\
%         \hline
%     \end{tabular}
%     \label{tab:perf_indicators}
% \end{table}

\begin{table}[h]
\scriptsize
    \centering
    \caption{Closed-loop performance comparison between the different control designs}
    \begin{tabular}{|c|c|c|}
        \hline
        Controller & Undershoot (\%)  & Settling Time $5\%$ (ms) \\
        \hline
        $C_{ZN}$ & 61  & 2.7 \\
        $C_{VRFT}$ & 60.8  & 1.8 \\
        $C_{VRFT_{AW}}$ & 11.4  & 0.9 \\
        \hline
    \end{tabular}
    \label{tab:perf_indicators}
\end{table}

\section{Conclusion} 
\pauline{This work details the data-driven control design of a power converter using VRFT \cite{ref5} and its anti-windup extension \cite{ref20}. On the presented example, a simple and reproducible choice of hyperparameters allow to obtain a PI controller, with or without anti-windup compensation, from a single set of input-output measurements. The performances are better than the considered baseline controller obtained through the Ziegler-Nichols tuning rules.}%These studies investigated the application of data-driven control techniques for power converters, focusing on their reliability and performance compared to classical methods such as PWM-based control and linearized model-based designs. A nonlinear state-space model of the OwnTech power converter was developed using state-space averaging. A PI controller is designed using the Ziegler-Nichols method. However, the Virtual Reference Feedback Tuning (VRFT) technique demonstrated better performance compared to the Ziegler-Nichols method. VRFT’s ability to estimate controllers quickly through least squares optimization makes it an effective choice for rapid controller design.\

% One significant challenge addressed in this work was integral wind-up in PI controllers, which often leads to instability due to power converter saturation. Anti-windup techniques were implemented to overcome this issue, resulting in stable and reliable control.
%  This research demonstrated the potential of data-driven control techniques, as it's ability of integrating the anti-windup phenomena in the estimation process makes preferable and faster.\\ 
 
Future work should focus on the experimental validation of the proposed method to confirm its practical effectiveness, in particular on more complex architecture. Additionally, investigating the performance of these techniques under more dynamic and challenging operating conditions \pauline{would be an interesting research path. Indeed, while this approach allows an easy and quick control tuning, there is no stability nor performance guarantees contrary to Lyapunov-based approaches. In the case of linear systems, this is enforced in data-driven control procedure through the choice of reference model, training dataset and controller structure. The underlying theoretical and methodological problem is then to extend this understanding to the case of nonlinear systems.}

% \subsection*{State Space Matrices:}

 % Right side for the figures

\end{document}